\documentstyle[aps,prb,twocolumn]{revtex}
\begin{document}
\draft
\title{Study of energy transfer in helium atom scattering from surfaces}
\author{A. \v{S}iber$^{a}$, B. Gumhalter$^{a,b}$ and J.P. Toennies$^{c}$}
\address{$^{a}$ Institute of Physics of the University, P.O. Box 304, 10001 Zagreb, Croatia \protect\\
$^{b}$ Abdus Salam International Centre for Theoretical Physics, I-34014 Trieste, Italy
\protect\\
$^{c}$ Max-Planck-Institut f\"{u}r Str\"{o}mungsforschung, D-37073 G\"{o}ttingen, Germany}
 
\maketitle

\begin{abstract}
Recently developed quantum mechanical theory of inelastic He atom scattering (HAS)
from solid surfaces is employed to analyze the energy
transfer between projectile particles (thermal energy He-atoms) and
vibrational degrees of freedom (phonons) characteristic of a variety of experimentally studied surfaces. We have first 
calculated the angular resolved energy transfer which can be
directly compared with the values deducible from the HAS time-of-flight spectra and a good agreement with  experimental data has been found. This enabled us to  calculate the total or angular
integrated energy transfer, which is of paramount importance in the studies of gas-surface scattering, but is neither  accessible in HAS (which yields only the angular resolved quantities), nor in the wind tunnel measurements for surfaces whose atomic composition and cleanliness must be maintained during the experiment. Here we present the results for prototype collision systems of this kind, viz. He$\rightarrow$Cu(001),  He $\rightarrow$Xe/Cu(111) and He$\rightarrow$Xe(111)
which are representative of the very
different types of surface vibrational dynamics and thereby provide an insight into some common properties of energy transfer in gas-surface scattering.

\end{abstract}

%-----------------------------------------------------------
%-----------------------------------------------------------
%\newpage

\section{Introduction}
\label{sec:intro}

For a long time the energy transfer in gas-surface collisions has been in the focus of interest of surface scientists. Special attention has been paid to the energy (heat) transfer in the free molecular flow regime because of its importance for understanding the flow interaction with space vehicles. The energy transfer in this regime of gas-surface collisions is  dominantly due to the gas interaction with vibrational degrees of freedom or phonons of the surface involved. The macroscopic features of heat transfer characteristic of the various technologically interesting surfaces are routinely investigated in wind tunnel experiments but the microscopic properties of the corresponding interactions can be assessed only by molecular beam scattering techniques. 

The major amount of information on the microscopic vibrational properties of a number of surfaces of both fundamental and practical interest has been acquired by utilizing the technique of mono-energetic He atom  scattering (HAS) combined with the energy resolution of the scattered beam by the time-of-flight (TOF) method (for a review see Ref. [\onlinecite{Hulpke}]).   
At present, this technique provides only the angular resolved quantities (e.g. energy transfer in angular resolved scattering events), and in order to obtain experimental information on the total energy transfer in gas-surface collisions the HAS studies must be complemented by wind tunnel measurements \cite{Legge}.
However, the studies of total energy transfer at surfaces whose structure, composition and cleanliness can be controlled only under the UHV conditions, are not possible in the wind tunnel environment. Hence, different methods for assessing the total energy transfer involving such surfaces must be employed in order to obtain the relevant information.  

In this work we demonstrate that investigations of vibrational properties of surfaces by the application of HAS-TOF technique also provide an excellent database for the assessment of total energy transfer in gas-surface scattering. Whereas the experimental data acquired in the highly quantum single phonon collision regime of HAS are primarily used to detect and identify the various surface localized vibrational modes, the available information from the single and multiphonon scattering regime combined with the quantum theory of HAS can be conveniently used to reveal the magnitudes of angular resolved and total energy transfer in gas-surface collisions. This is particularly important for a relatively large number of surfaces whose desired microscopic characteristics cannot be maintained in wind tunnel experiments which at present are the only method that can render the values of total energy transfer. 

In the following sections we first demonstrate how the quantum, microscopic theory of energy transfer in gas-surface collisions can be verified by comparing with the experimental data obtained from HAS-TOF measurements. Then we show how such a theory can be extended to calculations of the total energy exchange between the projectile particles and the target as a function of the surface temperature, projectile incident energy and angle, etc. The procedure is exemplified on selected prototype systems recently studied experimentally: clean  Cu(001) surface, monolayer of Xe atoms adsorbed on Cu(111) surface and the (111) surface of Xe multilayer crystal condensed on Pt(111). 
The properties of energy transfer involving  these surfaces with manifestly  different types of vibrational dynamics should be equally relevant to a variety of other systems investigated by HAS, as well as to the surfaces studied in wind tunnel experiments.

\section{Quantum description of atom-surface scattering and energy transfer}
\label{sec:theor}

The knowledge of the quantum scattering spectrum $N_{{\bf k_{i}},T_{s}}(\varepsilon)$ which gives 
the probability density that an atom with initial momentum 
$\hbar{\bf k_{i}}$ exchanges energy $\varepsilon$ with  
a surface at the temperature $T_{s}$ can be used to evaluate the total energy  
transfer $\mu$, which enters heat transfer and accommodation coefficients
\cite{Legge,Schaaf,Cerc}, according to the expression 
\begin{equation} 
\mu({\bf k_{i}},T_{s})=\int^{\infty}_{-\infty}\varepsilon N_{{\bf k_{i}},
T_{s}}(\varepsilon) d\varepsilon.
\label{eq:mu1}
\end{equation}
The quantity $N_{{\bf k_{i}},T_{s}}(\varepsilon)$, and thereby $\mu({\bf k_{i}},T_{s})$, 
is not directly accessible in typical HAS-TOF  
measurements from which most of the data is available at present. 
These experiments yield the {\em energy and angular resolved}
quantities usually only for fixed total 
scattering angle $\theta_{SD}=\theta_{i}+\theta_{f}$ where $\theta_{i}$ and $\theta_{f}$ denote the incident and final angle of scattering, respectively. The  TOF spectrum is directly proportional to the energy and 
lateral momentum  resolved scattering distribution 
$N_{{\bf k_{i}},T_{s}}(\varepsilon,\Delta{\bf K})$ (see Ref.
[\onlinecite{comment}]) in which $\varepsilon$ and $\Delta{\bf K}$ denote the amount of energy and lateral momentum exchanged between the particle and the surface. The values of $\varepsilon$ and $\Delta{\bf K}$ are connected through the conservation of total lateral momentum and energy in the course of collision, leaving $\varepsilon$ and $\theta_{f}$ as the only independent observation variables. 
Using this we can obtain the 
{\em angular resolved} energy transfer from:
\begin{equation}
\mu_{r}({\bf k_{i}},T_{s},\theta_{f})=
\frac{\int \varepsilon N_{{\bf k_{i}},T_{s}}(\varepsilon,\Delta{\bf
K}(\varepsilon))d\varepsilon}
{\int  N_{{\bf k_{i}},T_{s}}(\varepsilon,\Delta{\bf K}(\varepsilon))
d\varepsilon},
\label{eq:muresol}
\end{equation}
and the magnitude of $\mu_{r}({\bf k_{i}},T_{s},\theta_{f})$ 
 computed from the experimental
TOF spectra can be compared with the results based on theoretical $N_{{\bf k_{i}},T_{s}}(\varepsilon,\Delta{\bf K})$.  
The calculation of reliable multiphonon
$N_{{\bf k_{i}},T_{s}}(\varepsilon,\Delta{\bf K})$, from which 
$N_{{\bf k_{i}},T_{s}}(\varepsilon)$ is obtained by integration over 
$\Delta{\bf K}$, is difficult in the regimes in which multiphonon and quantum dynamics effects are 
important.  
However, recent progress in experiment and theory of HAS from surfaces 
makes it possible to measure and accurately calculate   
$N_{{\bf k_{i}},T_{s}}(\varepsilon,\Delta{\bf K})$ for some  prototype 
collision systems and thus obtain a reliable
estimate of the total energy transfer in these experiments.

In the regime of HAS in which the uncorrelated phonon processes are 
dominant the angular resolved scattering spectrum has the following unitary 
form  
\cite{BGL,HAS}:
\begin{eqnarray}
N_{{\bf k_{i}},T_{s}}(\varepsilon,\Delta{\bf K})
&=& 
\int^{\infty}_{-\infty} \frac{d\tau d^{2}{\bf R}}{(2\pi\hbar)^{3}}\nonumber\\ 
&\times&
e^{\frac{i}{\hbar}(\varepsilon\tau-\hbar(\Delta{\bf K}){\bf R})}
\exp[2W(\tau,{\bf R})-2W].
\label{eq:specEBA}
\end{eqnarray}
Here $2W(\tau,{\bf R})$ is the scattering function to be defined below,   
$2W=2W(0,0)$ gives the Debye-Waller exponent \cite{dwf} and $\tau$ and 
${\bf R}$ are auxiliary integration variables used to
project the states with $\varepsilon$ and $\Delta{\bf K}$ out of the integral
on the RHS of (\ref{eq:specEBA}). 
In the range of validity of  (\ref{eq:specEBA}) the energy and lateral 
momentum are conserved in each phonon exchange process and 
the values of $\varepsilon$ and $\Delta{\bf K}$ are constrained by the conservation of the total lateral momentum and energy in the collision.   
Using expression (\ref{eq:specEBA}) it is possible to write Eq. 
(\ref{eq:mu1}) as:
\begin{equation}
\mu({\bf k_{i}},T_{s})=i\frac{\partial}{\partial \tau} 
2W(\tau=0,{\bf R}=0),
\label{eq:mu1EBA}
\end{equation}
which can be readily calculated once  $2W(\tau,{\bf R})$ is established.
In the following the projectile-phonon coupling is assumed  linear in the adsorbate displacements since this gives the
dominant multiphonon contribution observed in HAS \cite{AM}, yielding \cite{HAS}:
\begin{eqnarray}
2W(\tau,{\bf R})&=&
\sum_{{\bf Q},j,k_{z}}
\left[ \mid {\cal V}^{{\bf K_{i},Q},j}_{k_{z},k_{zi}}(+)\mid^{2}[\bar{n}
(\omega_{{\bf Q}j})+1]  e^{-i(\omega_{{\bf Q}j}\tau -{\bf QR})}\right.\nonumber\\ 
&+&
\left. \mid {\cal V}^{{\bf K_{i},Q},j}_{k_{z},k_{zi}}(-) \mid^{2}\bar{n}
(\omega_{{\bf Q}j})  e^{i(\omega_{{\bf Q}j}\tau -{\bf QR})}\right].
\label{eq:WEBA}
\end{eqnarray}
Here ${\bf Q}$ and $j$ denote the lateral wave-vector and branch index of a 
normal phonon mode of frequency $\omega_{{\bf Q}j}$, respectively, 
${\bf k}=({\bf K},k_{z})$ where $k_{z}$ is the quantum number of 
distorted waves describing the projectile motion normal to the surface and  $\bar{n}(\omega_{{\bf Q}j})$ 
is the Bose-Einstein distribution.
The symbols $\mid{\cal V}^{{\bf
K_{i},Q},j}_{k_{z},k_{zi}}(\pm)\mid^{2}$ denote 
one phonon emission $(+)$ and  absorption $(-)$ scattering probabilities expressed in terms of the corresponding on-the-energy-shell transition matrix elements ${\cal V}^{{\bf
K_{i},Q},j}_{k_{z},k_{zi}}(\pm)$ normalized to unit particle current in the $z$-direction
\cite{HAS}. 
Substitution of (\ref{eq:WEBA}) into (\ref{eq:mu1EBA}) gives the
desired expression for the mean energy transfer in which
the so-called recoil term $\sum_{{\bf Q},j,k_{z}}\hbar\omega_{{\bf Q}j}[\mid {\cal V}^{{\bf K_{i},Q},j}_{k_{z},k_{zi}}(+) 
\mid^{2}-\mid {\cal V}^{{\bf K_{i},Q},j}_{k_{z},k_{zi}}(-) \mid^{2}]\bar{n}(\omega_{{\bf Q}j})$  determines the temperature dependence. 
Quite generally, the magnitude of the total energy transfer calculated from expression (\ref{eq:mu1EBA}) is much more sensitive to the character and variations of the polarization vectors of surface modes than to the changes in the functional behavior of their dispersion.   
A more detailed description of the scattering formalism used in the present study has been given in Ref. [\onlinecite{HAS}].

It should be pointed out that the above quoted expressions for calculating the scattering spectra and energy transfer are valid irrespective of the collision regime  (single versus multiphonon) they are applied to. Hence, the reliability of the scattering potentials and the description of the vibrational dynamics of the target employed in the model should first be tested by comparing the calculated and measured {\em angular resolved} TOF spectra or energy transfers in either the single or multiphonon scattering regime, or in both if the data are available, and then applied to the calculations of total energy transfer.

\section{Comparison of theoretical results with HAS-TOF data}

\subsection{He$\rightarrow$Cu(001) collisions}

The vibrational properties of the Cu(001) surface have been extensively studied by HAS \cite{Mason,Lapujoulade,pseudo,Cvetko,Hofmann} and these experiments have provided plentiful information on its vibrational dynamics whose characteristics are in many aspects also relevant to other clean monocrystal metal surfaces. This particularly applies to the pronounced intensities of the Rayleigh wave (RW) and the longitudinal resonance mode (LR) which dominate the spectra in the single phonon scattering regime. However, these specific features  of the single phonon scattering spectra which can be clearly traced across the first surface Brillouin zone (SBZ) are quickly washed out as the multiphonon scattering regime is approached. 
Here we shall focus on the He$\rightarrow$Cu(001) multiphonon scattering data reported in Ref. [\onlinecite{Hofmann}] because of their amenability to the analysis of the energy transfer. 
The analyses of these measurements \cite{comment} have shown that the multiphonon spectral intensities in the case of planar and atomically smooth surfaces are in the greatest part  determined by the coupling of scattered He atoms to surface vibrations with polarization dominantly perpendicular to the surface, in the present case to the Rayleigh phonons \cite{comment,Hofmann,Siber}. 
Hence, in our assessment of the energy transfer in HAS from Cu(001) surface we shall follow the description of the dynamic He-surface interaction outlined in Refs. [\onlinecite{comment,Siber}] with an improved modelling of the RW dispersion and polarization based on a dynamical matrix description used in connection with the Cu substrate \cite{PRL}, and of the static He-surface potential presented in Ref. [\onlinecite{Xenia}]. 

Following the approach outlined in Sec. \ref{sec:theor} and Refs. [\onlinecite{HAS,Xe}] we first calculate the multiphonon scattering spectra $N_{{\bf k_{i}},T_{s}}(\varepsilon,\Delta{\bf K})$ and compare them with the experimental HAS-TOF data \cite{Hofmann} to test the validity of the model potentials and vibrational dynamics parameters employed in the calculations. Then we use this as an input to calculate the angular resolved energy transfer (\ref{eq:muresol}) which can again be compared with the analogous quantity obtained from expression (\ref{eq:muresol}) by a direct integration of the corresponding experimental TOF spectra  and their first moments.

A comparison of the theoretical and experimental angular resolved energy transfer obtained from (\ref{eq:muresol}) by using expression (\ref{eq:specEBA}) and the datapoints of the measured scattering spectra, respectively, is given in Fig. \ref{Cu} as a function of the projectile angle of incidence and fixed incoming beam energy of 113 meV used in the experiments \cite{Hofmann}. The obtained agreement between experimental and theoretical results 
enables the application of the thus verified calculated quantities (\ref{eq:specEBA}) and (\ref{eq:WEBA}) in the evaluation of the total energy transfer using expression (\ref{eq:mu1EBA}). These calculations are reported in Sec. \ref{sec:mutot}.

\subsection{He$\rightarrow$Xe/Cu(111) collisions}

As the next typical example we consider the energy transfer to the 
adlayers of noble gas atoms adsorbed on metallic substrates because their vibrational properties have attracted considerable attention in the past two decades. The early interest in these systems was mainly due to the allegedly simple vibrational properties of such surfaces \cite{Sibener} which could be easier analyzed theoretically than the surfaces of pure metals, on the one hand, or of the metal compounds and ionic crystals, on the other hand. The later interest has been associated with the reduced effective dimensionality of adsorbed monolayers and the possibility to study phase transformations and thermodynamics of supported quasi-two-dimensional systems. The most recent interest is connected  with the development of the scanning tunneling microscope techniques which enable manipulations with adsorbates on the atomic scale.

A prototype system of a rare gas adlayer is the  commensurate $(\sqrt{3}\times\sqrt{3})R30^{0}$ monolayer of Xe atoms adsorbed on Cu(111) surface at substrate temperatures between $\sim$47 and $\sim$80 K \cite{PRL,Xe}. The adlayer vibrations in this system are almost completely decoupled from the vibrational dynamics of the underlying substrate \cite{PRL}, i.e. with a negligible admixture of the substrate modes and hence surface localized. 
Three Xe-adlayer induced modes or phonons are characterized according to their polarization properties. The first is a shear vertical or frustrated translation or S-mode with polarization perpendicular to the surface and very flat, practically negligible dispersion ($\omega_{S}(Q)=\omega_{0}=2.62$ meV). The second is the in-plane dominantly longitudinally polarized L-mode which strongly disperses over the first SBZ of the superstructure and exhibits a zone center gap of the order of 0.5 meV. The third is the dispersive transversely polarized or shear horizontal (SH) mode with the in-surface-plane polarization dominantly perpendicular to the phonon wavevector. 
This mode exhibits the same zone center gap but its frequency lies always below that of the L-mode. 
The only deviation from the described behaviors occurs in a very narrow range of the phonon wavevector $Q$ where the dispersion curves of these modes intersect (or rather exhibit an avoided crossing) with the dispersion curve of the RW of the Cu substrate. In this case their localization may extend over the first few substrate layers with the additional effect of mixed polarization of the mode \cite{PRL}. 
Outside the region of avoided crossing with the substrate RW these modes, besides being localized in the adlayer, are pure in character (negligible mixed polarizations) and their polarization eigenvectors are mutually orthogonal. 
Hence, this system lends itself as a particularly convenient  one for studying the energy transfer to pure surface localized vibrations.   

 The results of the calculations following the general procedure analogous to the one employed in the preceding subsection are depicted in Fig. \ref{Xe/Cu}. Here the angular resolved energy transfer is plotted as a function of the substrate temperature for several incoming He atom energies
and fixed experimental incident angle. We again find that the vibrations with dominant component of polarization perpendicular to the surface, here the S-modes, give preponderant contribution to the energy transfer.
The agreement between the thus obtained experimental and theoretical values is good, thereby providing a reliable starting point for the calculations of total energy transfer reported in the next section.

\subsection{He$\rightarrow$Xe(111) collisions}

As the third example we consider the energy transfer in He atom scattering from a van der Waals crystal, viz. the (111) surface of Xe multilayer condensed on Pt(111). Recent HAS angular distribution (diffraction) and TOF studies of this system \cite{Grahamprivate} reveal an ordered and almost defect-free surface sustaining several types of well resolved vibrational modes. This enables us to  apply the above described theoretical approach to this collision system as well. In particular, we were able to set up a dynamical matrix describing the vibrational properties of a Xe slab consisting of 60 layers and bounded by (111) surfaces by using the force constants calculated from the available Xe-Xe gas phase pair potentials \cite{Ramseyer}. The dispersion curves of the various phonon modes obtained from such a  dynamical matrix are in a fairly good agreement with the ones derived from the measured He$\rightarrow$Xe(111) TOF spectra \cite{Grahamprivate}. 

Application of expression (\ref{eq:muresol}) to the measured He$\rightarrow$Xe(111) scattering spectra \cite{Grahamprivate} yields the values of the angular resolved energy transfer shown in Fig. \ref{Xe111} as a function of the angle of incidence and fixed incoming He atom beam energy of 10 meV used in the experiments. We then calculated the angular resolved scattering spectrum (Eq. (3)) in the distorted wave Born approximation limit \cite{HAS} pertinent to the scattering regime in which the experimental data were recorded. This was used to obtain the corresponding theoretical energy transfer from Eq. (2) and the result, which is presented in Fig. 3, is consistent with the values deduced from the experimental spectra.
The general trends in the behavior of the energy transfer are similar to those characterizing the He$\rightarrow$Cu(001) collision system despite the large differences in the incident energy of the scattered He atoms and the substrate temperature.

\section{Calculations of the total energy transfer}
\label{sec:mutot}

A good agreement between the experimental and theoretical  angular resolved energy transfer in HAS established in the preceding section for three prototype collision systems enables us to employ the obtained results to estimate the total energy transfer in these experiments.
Figure \ref{mutot} illustrates the behavior of total energy transfer in the discussed systems as a function of the projectile incident energy over the interval which is relevant to HAS experiments as well as to the conditions of wind tunnel investigations of gas-surface interactions in space.  It is noticeable that the magnitude of the energy transfer is surface specific in the studied scattering regime, in contrast to the results from standard classical accommodation theories \cite{Legge,Schaaf}. In particular, 
the energy transfer is larger for surfaces with enhanced surface projected phonon density of low energy states with perpendicular polarization. Here it is largest for the Xe/Cu(111) surface because the corresponding S-phonon exhibits perpendicular polarization and Xe adlayer localization over almost the entire SBZ. 
This points to the dominant role which the surface modes with perpendicular polarization play in energy transfer in gas interactions with atomically smooth surfaces.

Consistent with these properties we also find that, in the scattering regime spanned by the collision parameters and energy interval of Fig. \ref{mutot}, the classical Baule formula for semiquantitative estimate of the energy transfer \cite{Trilling} applies only to the systems He$\rightarrow$Xe/Cu(111) and He$\rightarrow$Xe(111). 
In these two cases the frequencies of surface vibrations of Xe atoms (corresponding mode energies not exceeding 4 meV \cite{PRL,Grahamprivate}) are much smaller than the inverse collision time and hence the impulsive scattering limit in which the Baule formula holds is reached. This is also in accord with independent testings of the applicability of the Baule formula to He atom scattering from condensed multilayers of noble gas atoms on Si(100) surface \cite{Blahusch}. On the other hand, the surface Debye frequency of copper, $\omega_{D}^{Cu}$, is much higher ($\hbar\omega_{D}^{Cu}\sim$24 meV \cite{Hofmann}) and of the order of inverse collision time characteristic of the He$\rightarrow$Cu(001) system in the studied scattering regime. As a consequence, the impulsive scattering limit is not yet reached in the same energy interval which makes the Baule formula inapplicable.

\section{Conclusion}

We have presented a method for calculating the energy  transfer in the quantum regime of 
scattering of atoms from surfaces. The general features of the results obtained for three prototype gas-surface collision systems recently  studied by He atom scattering should be also relevant to a large variety of studies of free molecular flow past solid bodies performed in wind tunnel experiments.

\acknowledgments

 The work in Zagreb has been supported in part by the NSF grant JF-133. B.G. acknowledges the Senior Research Associateship of the Abdus Salam International Centre for Theoretical Physics in Trieste where a part of the theoretical work has been carried out. 
 
%\newpage

%\newpage

\begin{figure}
\caption{Comparison of the theoretical angular resolved energy transfer in He$\rightarrow$Cu(001) collisions (full line) calculated from expressions (\protect\ref{eq:specEBA}) and (\protect\ref{eq:WEBA}) with the values deduced from available experimental HAS-TOF spectra (open circles) using expression (\protect\ref{eq:muresol}), given as a function of $\Delta\theta=\theta_{i}-\theta_{SD}/2$
and for the scattering conditions as denoted. Inset: Comparison of the experimental multiphonon HAS-TOF (open circles) and theoretical (full curve) scattering spectra for $\Delta\theta$ corresponding to the experimental point denoted by arrow.} 
\label{Cu}
\end{figure}

\begin{figure}
\caption{Comparison of the theoretical angular resolved energy transfer in He$\rightarrow$Xe/Cu(111) collisions (full lines) calculated from expressions (\protect\ref{eq:specEBA}) and (\protect\ref{eq:WEBA}) with the values deduced from available experimental HAS-TOF spectra (open symbols) using expression (\protect\ref{eq:muresol}), given as a function of the substrate temperature $T_{s}$ for the scattering conditions as denoted. Inset: Comparison of experimental multiphonon HAS-TOF (open circles) and theoretical (full curve) scattering spectrum corresponding to the scattering conditions of the experimental point denoted by arrow.}
\label{Xe/Cu}
\end{figure} 

\begin{figure}
\caption{Comparison of the theoretical angular resolved energy transfer in He$\rightarrow$Xe(111) collisions (full line) calculated from expressions (\protect\ref{eq:specEBA}) and (\protect\ref{eq:WEBA}) with the values deduced from available experimental HAS-TOF spectra (squares) using expression (\protect\ref{eq:muresol}), given as a function of  $\Delta\theta=\theta_{i}-\theta_{SD}/2$ and for the scattering conditions as denoted. Inset: Comparison of experimental HAS-TOF (open squares) and theoretical (full curve) scattering spectrum for $\Delta\theta$ corresponding to the experimental point denoted by arrow.}
\label{Xe111}
\end{figure}

\begin{figure}
\caption{Total energy transfer characteristic of He$\rightarrow$Cu(001), He$\rightarrow$Xe/Cu(111) and He$\rightarrow$Xe(111) collisions calculated from expression (\protect\ref{eq:mu1EBA}) and plotted as a function of the incoming He beam energy for fixed incident angle and substrate temperature. Starred line is the energy transfer obtained from the classical Baule formula applied to He$\rightarrow$Xe binary collisions.}
\label{mutot}
\end{figure}

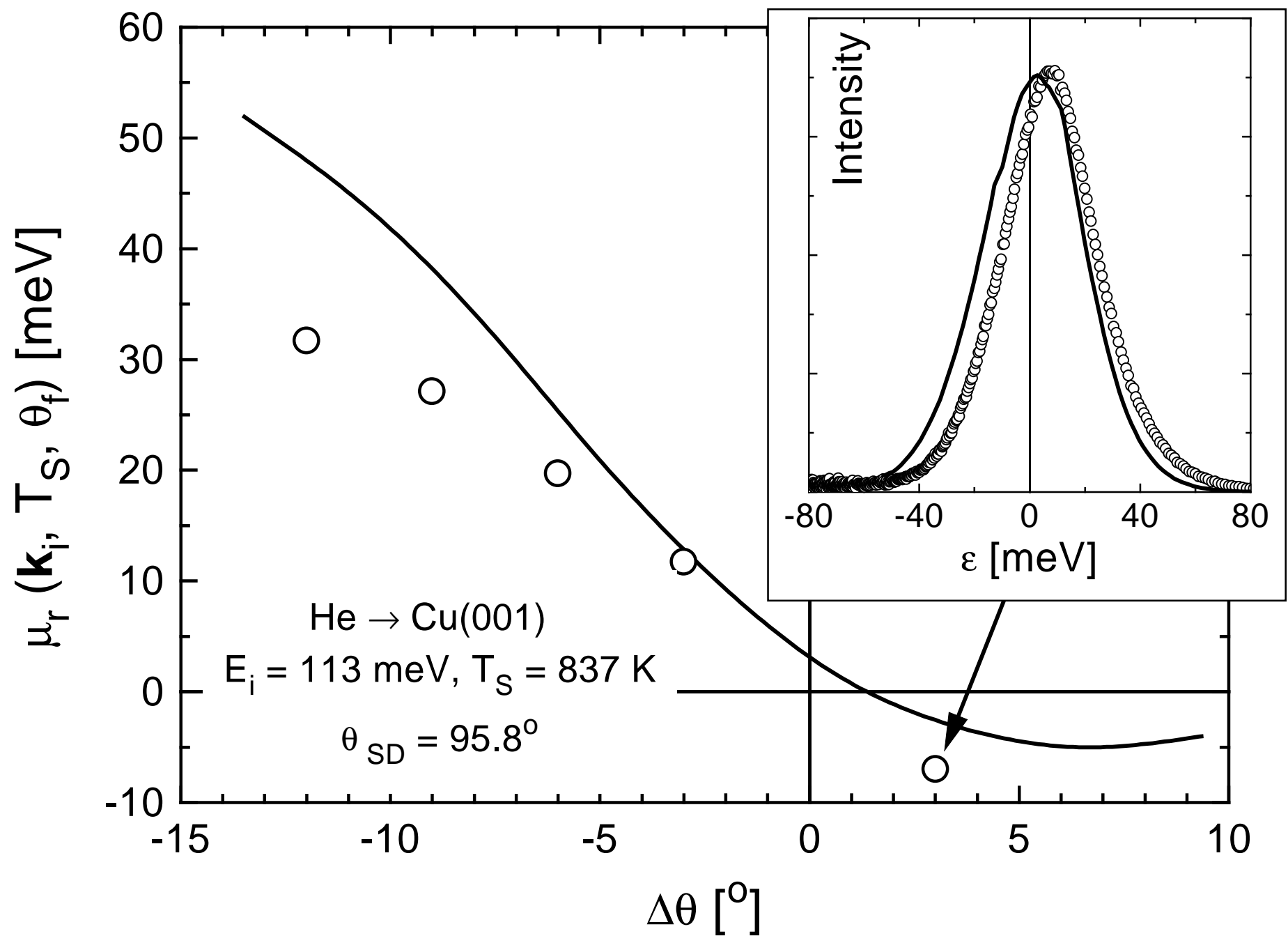
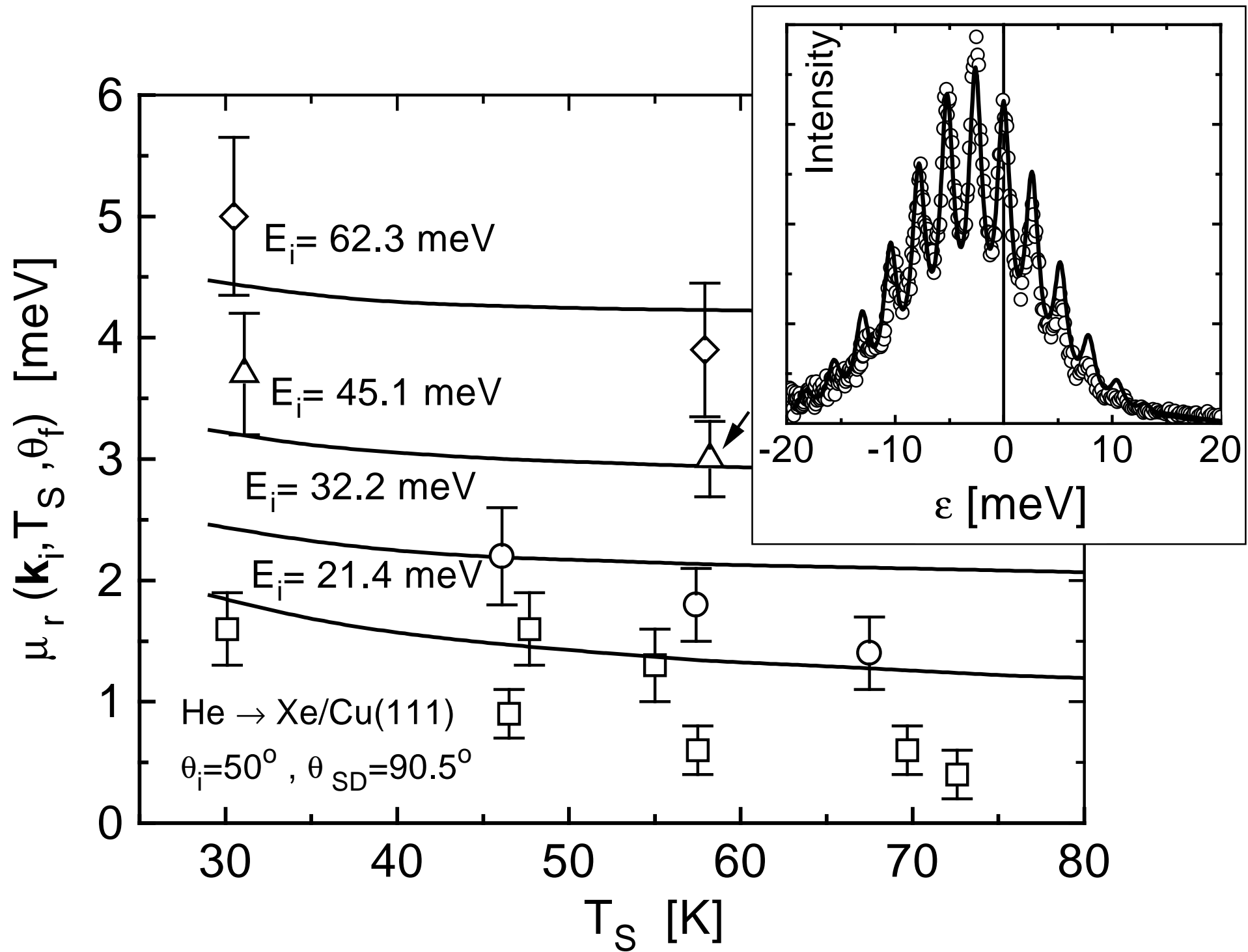
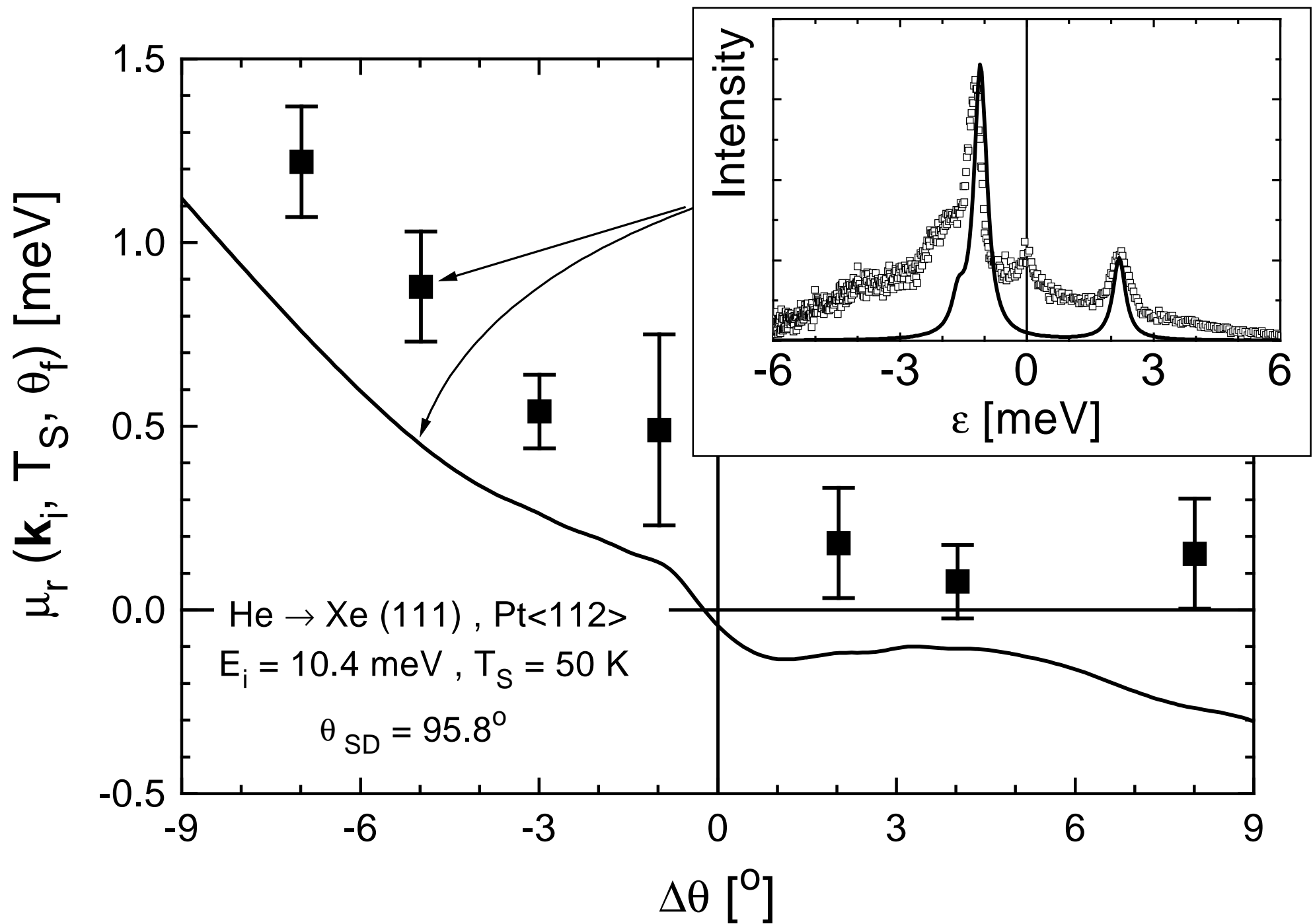
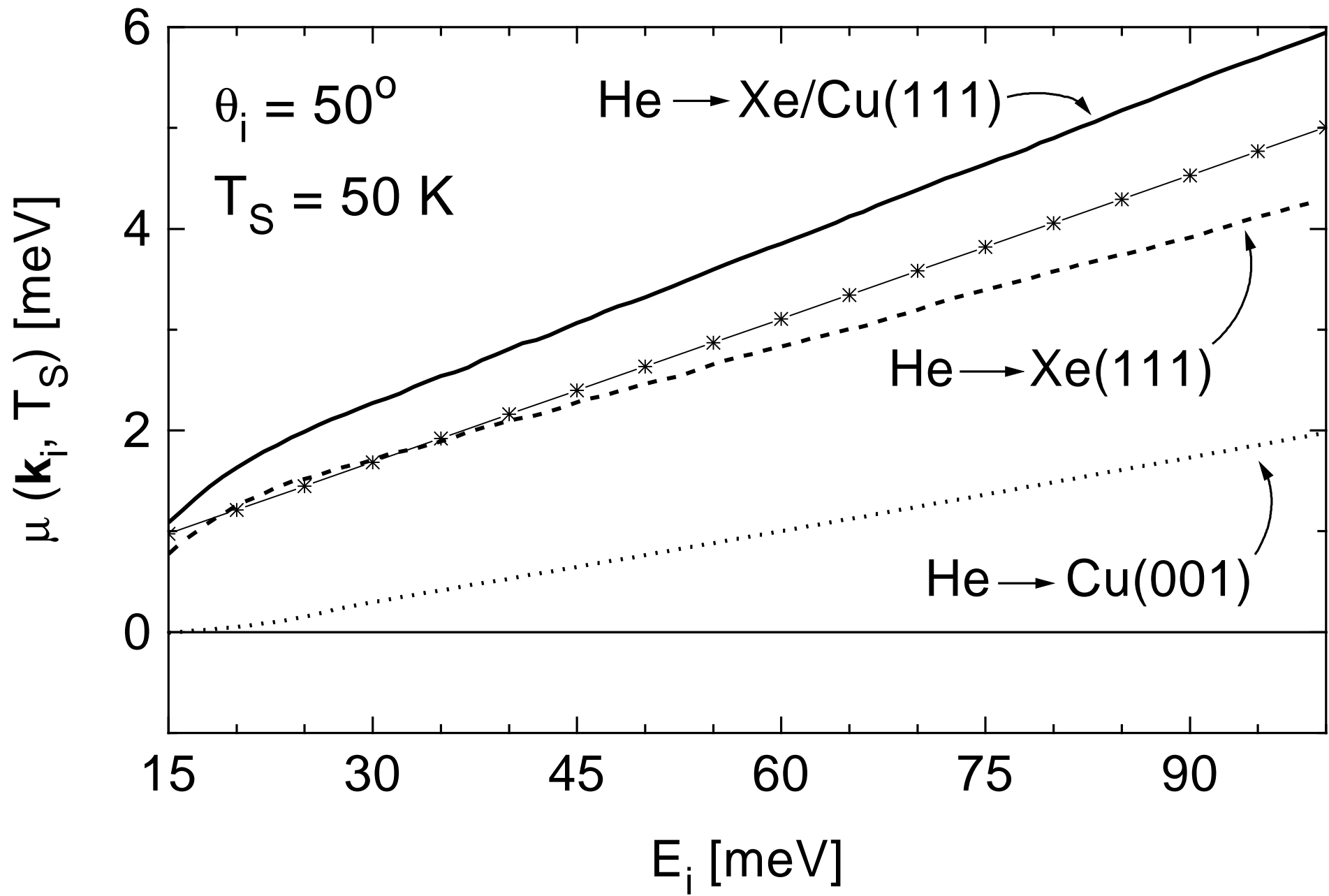
\end{document}